\documentclass[a4paper,11pt]{article}

\usepackage{amssymb,amsmath,amsthm}
\usepackage{texdraw}
\usepackage{epic,eepic}
\usepackage{authblk}
\usepackage{breqn}

\newtheorem{thm}{Theorem}[section]

\theoremstyle{remark}
\newtheorem{rem}[thm]{Remark}

\begin{document}

\title{\textbf{Symmetries of charged particle motion\\under time-independent electromagnetic fields}}
\author{Nikos Kallinikos and Efthymia Meletlidou}
\affil{Department of Physics, Aristotle University of Thessaloniki, GR-54124 Thessaloniki, Greece
\tt{kallinikos@auth.gr, efthymia@auth.gr}}
\date{\today}

\maketitle

\begin{abstract}
A symmetry analysis is presented for the three-dimensional nonrelativistic motion of charged particles in arbitrary stationary electromagnetic fields. The general form of the Lie point symmetries is found along with the fields that respect them, considering non-trivial cases of physical interest. The restrictions placed upon the electromagnetic field yield five classes of solutions, expressed in terms of the vector and scalar potentials. The Noether type symmetries are also investigated and their corresponding invariants are found. A second integral of motion, besides the Hamiltonian, results in three general cases. Finally, a relation between the symmetries of the charged particle motion and the symmetries of the magnetic field lines is established. 
\end{abstract}

\section{Introduction}

Lie symmetry methods are one of the few systematic tools for studying nonlinear differential equations, either ordinary or partial, revealing, if any, their integrability properties. In dynamical systems, continuous symmetries can be used to reduce the order of the system and in some cases even completely integrate it. Along with Noether's theorem for variational problems, they may also yield conservation laws, which restrict solutions to an invariant manifold. Therefore, the direct search for symmetries in certain systems has received a great deal of attention over the past few decades. Another usage of symmetry analysis, which has been adopted more and more often, is to classify all possible symmetry groups admitted by a wide family of differential equations. Such results determine under which conditions a system of general form may possess one or more symmetries.

Following this approach, we investigate the symmetry properties of the Newtonian motion of a charged particle in an arbitrary, yet stationary, electromagnetic field. Obeying the Lorentz force law, the system is expressed by three second-order, autonomous differential equations, which in vector form are
\begin{equation}\label{lor}
\ddot{\boldsymbol x}=\dot{\boldsymbol x}\times\boldsymbol B(\boldsymbol x)+\boldsymbol E(\boldsymbol x),
\end{equation}
where $\boldsymbol B$ and $\boldsymbol E$ are smooth vector functions of the position $\boldsymbol x$ alone, considered in the Euclidean space $\mathbb{R}^3$, representing the magnetic and electric fields, respectively, while the dot stands for derivation with respect to time $t$. 
For $\boldsymbol{B}=\mathbf{0}$, the above system reduces to a general problem of Classical Mechanics, for which a complete symmetry group classification can be found in \cite{dam-sop2,tsa-pa1} in two dimensions and \cite{dam-sop3a,dam-sop3b,tsa-pa2} in three. A symmetry analysis for the two-dimensional case of system (\ref{lor}), where the magnetic field has a constant direction in space, has also been made \cite{ha-go1,ha-go2}, even for time-dependent electromagnetic fields. Therefore this case, along with trivial ones of less physical interest, where either one of the fields is homogeneous, will not be considered here. Our intention is to cover a variety of applications, in which the electromagnetic field may be quite complicated, such as those occuring in plasma physics and fusion devices. A series of simpler problems for particular choices of the functions $\boldsymbol{B}$ and $\boldsymbol{E}$ is presented in \cite{rit}.

In the present work, we search for Lie as well as Noether point symmetries admitted by the above equations, leaving the electromagnetic field as arbitrary as possible, neglecting the previously mentioned cases. We find that the general form of Lie point symmetries is a combination of dilations, rotations and translations, i.e. they are linear. Starting with an unprescribed electromagnetic field, the symmetry condition, apart from the symmetry generator, eventually places restrictions on the functions $\boldsymbol B$ and $\boldsymbol E$, too. The latter are expressed through three coupled first order linear partial differential equations for each field. Taking into account Maxwell's equations as well, these equations are solved in terms of the vector and scalar potentials, $\boldsymbol{A}$ and $\Phi$. Thus, we determine the general form of electromagnetic fields of physical interest, for which systems of class (\ref{lor}) admit point symmetries, along with the general form of the symmetries themselves. More precisely we obtain families of potentials that are compatible with a second Lie point symmetry, besides the expected time translations. Since, in terms of $\boldsymbol{A}$ and $\Phi$, the problem becomes a variational one, we also check which of the symmetries found are of Noether type. It turns out that, with the exception of one dilation, all other symmetries are variational, too. We find the corresponding invariants, focusing on integrals of motion, over and beyond the well-known Hamiltonian of the system. An important aspect is that these symmetries are also preserved by the magnetic (and the electric) field itself, and the induced integral for the magnetic field lines is also found.

This paper is organised as follows. In section 2, we determine the most general form of the Lie point symmetry generator of equations (\ref{lor}), along with the constraints of the electromagnetic field. These are tackled in the next section, where Maxwell's equations are also imposed for consistency with any real physical problem. Consequently, the previous conditions on $\boldsymbol B$ and $\boldsymbol E$ are expressed in terms of $\boldsymbol{A}$ and $\Phi$, and then solved. We obtain five classes of solutions, each of which corresponds to one specific symmetry, apart from time translations. In section 4, the general form of Noether point symmetries and the related electromagnetic fields are also found. The corresponding integrals of motion are derived in the following section. The existence of a second constant of motion, functionally independent of the Hamiltonian of the system, is considered, and three types of such integrals are obtained. Finally, section 6 introduces a comparison of the charged particle motion with the dynamics of the magnetic field itself, where the two systems are tested for common symmetries. An integral of motion for the magnetic field lines is obtained, as well.

\section{Lie point symmetries}
\label{sec2}
Using a Cartesian frame of reference, system (\ref{lor}) can also be written as
\begin{equation}
\label{sys}
\ddot{x}_i=\epsilon_{ijk}\dot{x}_j\!B_k+E_i,
\end{equation}
where $\epsilon_{ijk}$ is the Levi-Civita symbol. Einstein's summation convention has been adopted, assuming all indices from now on and throughout the rest of this paper take values from 1 to 3.

We consider infinitesimal point transformations of the form
\begin{align}
\label{trans1}
\widetilde{t}&=t+\epsilon\xi(t,\boldsymbol x)+O\!\left(\epsilon^2\right),\\
\label{trans2}
\widetilde{x}_i&=x_i+\epsilon\varphi_i(t,\boldsymbol x)+O\!\left(\epsilon^2\right),
\end{align}
generated by the vector field
\begin{equation}
\label{X}
X=\xi\frac{\partial}{\partial t}+\varphi_i\frac{\partial}{\partial x_i}.
\end{equation}
The symmetry condition then reads
\begin{equation}
\label{sc}
X^{(2)}\!\left(\ddot{x}_i-\epsilon_{ijk}\dot{x}_j\!B_k-E_i\right)=0,
\end{equation}
whenever equations (\ref{sys}) hold, where $X^{(2)}$ represents the second prolongation of $X$. For a detailed description of symmetry theory in differential equations, see \cite{bluman,olver,stephani}.

After substituting the second derivatives, equations (\ref{sc}) take the form of polynomials of third degree in terms of the first derivatives $\dot{x}_i$. Being identities for all $t, x_i$ and $\dot{x}_i$, they finally break up into a larger set of partial differential equations, the so-called determining equations,
\begin{align}
\label{d1}
\xi_{x_jx_k}&=0,\\
\label{d2}
2\varphi_{i_{x_jx_k}}-\epsilon_{ijl}\xi_{x_k}B_l+\epsilon_{ilk}\xi_{x_j}B_l-\delta_{ij}\!\left(2\xi_{tx_k}-\epsilon_{lmk}\xi_{x_l}B_m\right)-\delta_{ik}\!\left(2\xi_{tx_j}+\epsilon_{ljm}\xi_{x_l}B_m\right)&=0,\\
\label{d3}
2\varphi_{i_{tx^j}}+\epsilon_{ljk}\varphi_{i_{x_l}}B_k-\epsilon_{ijk}\xi_tB_k-2\xi_{x_j}E_i-\epsilon_{ilk}\varphi_{l_{x_j}}B_k-\epsilon_{ijk}\varphi_lB_{k_{x_l}}-\delta_{ij}\!\left(\xi_{tt}+\xi_{x_l}E_l\right)&=0,\\
\label{d4}
\varphi_{i_{tt}}+\varphi_{i_{x_j}}E_j-2\xi_tE_i-\epsilon_{ijk}\varphi_{j_t}B_k-\varphi_jE_{i_{x_j}}&=0,
\end{align}
where $\delta_{ij}$ is Kronecker's delta and subscripts $t$ or $x_i$ denote partial differentiation with respect to each of them.

Starting from the top, the first set of equations yields $\xi(t,\boldsymbol x)=f_i(t)x_i+w(t)$. As can be seen from the second one, mixed second-order partial derivatives $\varphi_{i_{x_jx_k}}$ and $\varphi_{i_{x_kx_j}}$ are already considered equal. Imposing further integrability conditions on (\ref{d2}), i.e. $\varphi_{l_{x_ix_jx_k}}=\varphi_{l_{x_ix_kx_j}}=\varphi_{l_{x_kx_jx_i}}$, we find, after some investigation, that $f_i=0$ for non-constant $\boldsymbol B$. This in turn means that
\begin{align}
\label{g1}
\xi(t,\boldsymbol x)&=w(t),\\
\label{g2}
\varphi_i(t,\boldsymbol x)&=h_{ij}(t)x_j+h_i(t)
\end{align}
where $w,h_{ij},h_i$ are arbitrary functions of $t$.

After substitution of the above expressions, equations (\ref{d3}) for $i=j$ take the form
\begin{align}
\begin{split}
\label{d3a}
2\dot{h}_{11}+\left(h_{13}+h_{31}\right)B_2-\left(h_{12}+h_{21}\right)B_3-\ddot{w}&=0,\\
2\dot{h}_{22}+\left(h_{12}+h_{21}\right)B_3-\left(h_{23}+h_{32}\right)B_1-\ddot{w}&=0,\\
2\dot{h}_{33}+\left(h_{23}+h_{32}\right)B_1-\left(h_{13}+h_{31}\right)B_2-\ddot{w}&=0,
\end{split}
\end{align}
from which follows the relation
\begin{equation}
\label{g3}
2h_{ii}=3\dot{w}+3c,
\end{equation}
where $c$ is a constant.

The rest of (\ref{d3}) for $i\neq j$ can be rearranged into two sets, by taking $ij$- and $ji$-equations in pairs. Adding the two equations of each pair gives the first set,
\begin{align}
\begin{split}
\label{d3b}
2\left(\dot{h}_{12}+\dot{h}_{21}\right)-\left(h_{13}+h_{31}\right)B_1+\left(h_{23}+h_{32}\right)B_2+2\left(h_{11}-h_{22}\right)B_3&=0,\\
2\left(\dot{h}_{13}+\dot{h}_{31}\right)+\left(h_{12}+h_{21}\right)B_1+2\left(h_{33}-h_{11}\right)B_2-\left(h_{23}+h_{32}\right)B_3&=0,\\
2\left(\dot{h}_{23}+\dot{h}_{32}\right)+2\left(h_{22}-h_{33}\right)B_1-\left(h_{12}+h_{21}\right)B_2+\left(h_{13}+h_{31}\right)B_3&=0.
\end{split}
\end{align}

Before proceeding with the second one, direct inspection of systems (\ref{d3a}) and (\ref{d3b}) shows that unless
\begin{equation}
\label{g4}
h_{ij}=-h_{ji},~~i\neq j
\end{equation}
hold, then the magnetic field is at best of constant direction. This case has already been investigated in \cite{ha-go2}.

So, in light of (\ref{g4}), equations (\ref{d3b}) trivially result in
\begin{equation}
\label{g5}
h_{11}=h_{22}=h_{33},
\end{equation}
while the remaining of equations (\ref{d3}), taking also into account (\ref{g3}), yield
\begin{equation}
\label{Bt}
\left(h_{jk}x_k+h_j\right)\frac{\partial B_i}{\partial x_j}=\left(c-h_{jj}\right)\!B_i+h_{ij}B_j+\epsilon_{ijk}\dot{h}_{jk},
\end{equation}
placing inevitably restrictions upon $\boldsymbol B$.

The latter show that, for a time-independent magnetic field, the functions $h_{ij},h_i$ must be constants. This can be seen, if we differentiate the last three equations with respect to $x_i$, and then treat $h_{ij},h_i$ and $c$ as the unknowns. Hence, we arrive at an algebraic system of nine equations in eight unknowns (considering (\ref{g4})-(\ref{g5})), which is also linear and homogeneous. In the generic case, the rank of this system is 7, and therefore we can solve for $h_{ij}$ and $h_i$ in terms of $c$, which is an absolute constant, and the derivatives of the magnetic field. Since $\boldsymbol B$ does not depend explicitly on time, all these solutions have to be constants.

Thus, equations (\ref{Bt}) take the form
\begin{equation}
\label{B}
\left(h_{jk}x_k+h_j\right)\frac{\partial B_i}{\partial x_j}=\left(c-h_{jj}\right)\!B_i+h_{ij}B_j,
\end{equation}
while (\ref{g3}), with regard to (\ref{g5}), can also be integrated, arriving at a linear expression in $t$ for $\xi$,
\begin{equation}
\label{g6}
w(t)=\left(2h_{11}-c\right)t+h_0,
\end{equation}
where $h_0$ is an arbitrary constant. Finally, the last set of the determining equations after substitution yield
\begin{equation}
\label{E}
\left(h_{jk}x_k+h_j\right)\frac{\partial E_i}{\partial x_j}=\left(2c-\frac{4}{3}h_{jj}\right)\!E_i+h_{ij}E_j,
\end{equation}
expressing the constraints on the electric field.

To summarize, from the relations (\ref{g1}), (\ref{g2}), (\ref{g4}), (\ref{g5}) and (\ref{g6}), the most general form of the symmetry generator is found to be
\begin{align}
\begin{split}
\label{sym0}
\xi&=\left(2h_{11}-c\right)t+h_0,\\
\varphi_1&=h_{11}x+h_{12}y-h_{31}z+h_1,\\
\varphi_2&=-h_{12}x+h_{11}y+h_{23}z+h_2,\\
\varphi_3&=h_{31}x-h_{23}y+h_{11}z+h_3.
\end{split}
\end{align}
We should note, however, that while all other constants are related through (\ref{B}) and (\ref{E}), $h_0$ is completely arbitrary. 
Thus, the symmetry algebra is spanned by $\partial_t$ and a second symmetry of the general form
\begin{align}
\label{sym1}
\text{v}_{\text{L}}=\left(2h_{11}-c\right)t\frac{\partial}{\partial t}+\varphi_1\frac{\partial}{\partial x}+
\varphi_2\frac{\partial}{\partial y}+\varphi_3\frac{\partial}{\partial z},
\end{align}
which is actually a sum of dilations, rotations and translations
\begin{align}
\label{sym2}
\nonumber\text{v}_{\text{L}}=\,&h_{11}\!\left(2t\frac{\partial}{\partial t}+x\frac{\partial}{\partial x}+y\frac{\partial}{\partial y}+z\frac{\partial}{\partial z}\right)-ct\frac{\partial}{\partial t}\,+\\
\nonumber&h_{23}\!\left(z\frac{\partial}{\partial y}-y\frac{\partial}{\partial z}\right)+h_{31}\!\left(x\frac{\partial}{\partial z}-z\frac{\partial}{\partial x}\right)+h_{12}\!\left(y\frac{\partial}{\partial x}-x\frac{\partial}{\partial y}\right)+\\
&h_1\frac{\partial}{\partial x}+h_2\frac{\partial}{\partial y}+h_3\frac{\partial}{\partial z}.
\end{align}  
While the former, expected for every autonomous system, appears whatever the (stationary) fields $\boldsymbol{B}$ and $\boldsymbol{E}$ may be, the latter is admitted only when $\boldsymbol{B}$ and $\boldsymbol{E}$ satisfy equations (\ref{B}) and (\ref{E}).

So, leaving the anticipated $\partial_t$ aside, we focus on the existence of the general symmetry $\text{v}_{\text{L}}$. The next task to complete the symmetry analysis is to find the form of the electromagnetic field, respecting the symmetry condition, i.e. solve equations (\ref{B}) and (\ref{E}).

\section{The form of the Electromagnetic field}
The solutions to (\ref{B}) and (\ref{E}) describe the general form the vector functions $\boldsymbol{B}$ and $\boldsymbol{E}$ must have in order for system (\ref{lor}) to admit symmetry (\ref{sym1}). In physics, however, the electromagnetic field obeys certain laws, which altogether constitute another set of equations, namely Maxwell's equations. Gauss's and Ampere's laws simply determine the source of the electric and magnetic fields, respectively, i.e. the charge and current densities. But the rest two represent conditions that $\boldsymbol{B}$ and $\boldsymbol{E}$ have to satisfy, and are inextricably connected with every real problem (\ref{lor}). In the case of stationary fields, these are the homogeneous equations
\begin{align}
\label{maxB}
\nabla\cdot\boldsymbol{B}&=0,\\
\label{maxE}
\nabla\times\boldsymbol{E}&=0.
\end{align}
Instead of imposing these conditions, we recall their basic consequence, that the magnetic field can be derived from a vector potential $\boldsymbol{A}=(A_1,A_2,A_3)$, while the electric field from a scalar one $\Phi$,
\begin{align}
\label{maxA}
\boldsymbol{B}&=\nabla\times\boldsymbol{A},\\
\label{maxPh}
\boldsymbol{E}&=-\nabla\Phi.
\end{align}

Replacing the solutions (\ref{maxA})-(\ref{maxPh}) in (\ref{B}) and (\ref{E}) provides us with the restrictions that now $\boldsymbol{A}$ and $\Phi$ must satisfy in order to respect the Lie point symmetry condition. The equations we end up with, after integration, can be cast into the following form:
\begin{align}
\label{AF}
\left(h_{jk}x_k+h_j\right)\frac{\partial A_i}{\partial x_j}&=\left(c-\frac{2}{3}h_{jj}\right)\!A_i+h_{ij}A_j+\frac{\partial f}{\partial x_i},\\
\label{Ph}
\left(h_{jk}x_k+h_j\right)\frac{\partial\Phi}{\partial x_j}&=2\left(c-\frac{1}{3}h_{jj}\right)\!\Phi-k,
\end{align}
where $f$ is an arbitrary function of $\boldsymbol{x}$ and $k$ some constant. Furthermore, the gauge invariance of the vector potential, $\boldsymbol{A}\longrightarrow\boldsymbol{A}+\nabla g$, admitted by (\ref{maxB}), can, in fact, guarantee the existence of an equivalent $\boldsymbol{A}$, such that
\begin{equation}
\label{A}
\left(h_{jk}x_k+h_j\right)\frac{\partial A_i}{\partial x_j}=\left(c-\frac{2}{3}h_{jj}\right)\!A_i+h_{ij}A_j,
\end{equation}
for $g$ satisfying
\begin{equation}
\label{g}
\boldsymbol{\varphi}\cdot\nabla g-cg=f,
\end{equation}
where $\boldsymbol{\varphi}$ is the vector with entries $\varphi_i$. Thus, in case of true electromagnetic fields, that certainly have to comply with Maxwell's equations, the solutions to (\ref{Ph})-(\ref{A}) describe through (\ref{maxA}) and (\ref{maxPh}) the ones for which system (\ref{lor}) has the symmetry (\ref{sym1}).

Consequently, we can either treat $\boldsymbol{B}$ and $\boldsymbol{E}$ only as functions entering the system, or we can view them as part of a bigger physical problem that also includes Maxwell's equations. 
In what follows, we focus on that second case and give the solutions $\boldsymbol{A}$ and $\Phi$ of (\ref{Ph})-(\ref{A}). For reasons that will be apparent, the potentials that the electromagnetic field comes from, besides carrying more information, are more convenient to use in this situation. The form of $\boldsymbol{B}$ and $\boldsymbol{E}$ can then be found through (\ref{maxA})-(\ref{maxPh}). If, however, one wishes to determine them without taking into account Maxwell's equations, the task would be similar. Equations (\ref{B}) and (\ref{E}) can be solved independently of each other, and obviously by the same means as (\ref{A}) can. Either one of them is a system of three coupled first-order linear partial differential equations, and in the case of the vector potential its explicit form is
\begin{align}
\begin{split}
\label{Ai}
\varphi_1(x,y,z)\frac{\partial A_1}{\partial x}+\varphi_2(x,y,z)\frac{\partial A_1}{\partial y}+\varphi_3(x,y,z)\frac{\partial A_1}{\partial z}&=\left(c-h_{11}\right)A_1+h_{12}A_2-h_{31}A_3,\\
\varphi_1(x,y,z)\frac{\partial A_2}{\partial x}+\varphi_2(x,y,z)\frac{\partial A_2}{\partial y}+\varphi_3(x,y,z)\frac{\partial A_2}{\partial z}&=-h_{12}A_1+\left(c-h_{11}\right)A_2+h_{23}A_3,\\
\varphi_1(x,y,z)\frac{\partial A_3}{\partial x}+\varphi_2(x,y,z)\frac{\partial A_3}{\partial y}+\varphi_3(x,y,z)\frac{\partial A_3}{\partial z}&=h_{31}A_1-h_{23}A_2+\left(c-h_{11}\right)A_3.\\
\end{split}
\end{align}
The other two have almost the exact same form, differing only by a minor factor. Thus, we present in detail a way of solving (\ref{Ai}), and then (\ref{B}) and (\ref{E}) can be treated accordingly.

First of all we want to uncouple the system, which in vector form may be written as
\begin{equation}
\label{AM}
(\boldsymbol{\varphi}\cdot\nabla)\boldsymbol{A}=Q\boldsymbol{A},
\end{equation}
where $Q$ is the $3\times3$ square matrix
\begin{equation}
Q=\left(\begin{matrix}
c-h_{11}&h_{12}&-h_{31}\\
-h_{12}&c-h_{11}&h_{23}\\
h_{31}&-h_{23}&c-h_{11}
\end{matrix}\right).
\end{equation}
The form of the equations allows one to do so, when $Q$ is diagonalizable. The above matrix however is not, since it has two complex eigenvalues, $\lambda_{1,2}=c-h_{11}\pm ih$, and only one real, $\lambda_3=c-h_{11}$, where $h=\left(h_{12}^2+h_{31}^2+h_{23}^2\right)^{\frac{1}{2}}$. Nevertheless, we can still separate one equation from the other two, by setting $\boldsymbol{A}=P\boldsymbol{\bar{A}}$, where $P=(\begin{matrix}u&v&e\end{matrix})$ is the matrix of the eigenvectors $u\pm iv,e$ of $Q$ corresponding to the eigenvalues $\lambda_{1,2},\lambda_3$, respectively. Then, left multiplication of equation (\ref{AM}) with the inverse of $P$ leads to
\begin{equation}
\label{AbM}
(\boldsymbol{\varphi}\cdot\nabla)\boldsymbol{\bar{A}}=\bar{Q}\boldsymbol{\bar{A}},
\end{equation}
where
\begin{equation}
\bar{Q}=P^{-1}QP=
\left(\begin{matrix}
c-h_{11}&h&0\\
-h&c-h_{11}&0\\
0&0&c-h_{11}
\end{matrix}\right).
\end{equation}
Thus, the third of equations (\ref{AbM}) for $\bar{A}_3$ is detached from the others. To uncouple the rest two we need a nonlinear transformation. Their form, as it can be seen from the matrix $\bar{Q}$, naturally implies taking polar coordinates in the $\bar{A}_1\bar{A}_2$ plane.

In conclusion, the following transformation in the dependent variables, i.e. the components of the vector potential,
\begin{align}
\begin{split}
\label{tranA}
a_1&=\sqrt{\bar{A}_1^2+\bar{A}_2^2}\\
a_2&=\arctan\!\left(\frac{\bar{A}_2}{\bar{A}_1}\right)~~~~~\text{and}~~~~\boldsymbol{\bar{A}}=P^{-1}\boldsymbol{A},\\
a_3&=\bar{A}_3
\end{split}
\end{align}
where $P$ is defined through the eigenvectors of $Q$, uncouples the three equations of (\ref{Ai}). Indeed, the latter take the form
\begin{align}
\begin{split}
\label{ai}
\varphi_1(x,y,z)\frac{\partial a_1}{\partial x}+\varphi_2(x,y,z)\frac{\partial a_1}{\partial y}+\varphi_3(x,y,z)\frac{\partial a_1}{\partial z}&=\left(c-h_{11}\right)a_1,\\
\varphi_1(x,y,z)\frac{\partial a_2}{\partial x}+\varphi_2(x,y,z)\frac{\partial a_2}{\partial y}+\varphi_3(x,y,z)\frac{\partial a_2}{\partial z}&=-h,\\
\varphi_1(x,y,z)\frac{\partial a_3}{\partial x}+\varphi_2(x,y,z)\frac{\partial a_3}{\partial y}+\varphi_3(x,y,z)\frac{\partial a_3}{\partial z}&=\left(c-h_{11}\right)a_3.\\
\end{split}
\end{align}
Now, each equation can be solved independently by the method of characteristics. Even more conveniently, all three of them are essentially (the first and the third exactly) the same. 

This means that three out of four characteristic equations for the above equations are common in each case, forming, in light of (\ref{sym0}), a linear dynamical system,
\begin{equation}
\label{char-eq}
\frac{d\boldsymbol{x}}{ds}=H\boldsymbol{x}+\boldsymbol{h},
\end{equation}
where $H$ is the $3\times3$ square matrix $(h_{ij})$ and $\boldsymbol{h}$ the column vector $\left(h_1,h_2,h_3\right)$. Since the matrices $Q$ and $H$ share the same eigenvectors, the homogeneous counterpart of the above equations can be easily resolved, very similarly to (\ref{AM}). Only now, due to the term $h_i$, we need to make a slight adjustment. More specifically, by introducing the new independent variables,
\begin{align}
\begin{split}
\label{tranx}
\widetilde{x}&=\sqrt{\bar{x}^2+\bar{y}^2}\\
\widetilde{y}&=\arctan\!\left(\frac{\bar{y}}{\bar{x}}\right)~~~~\text{and}~~~~\boldsymbol{\bar{x}}=P^{-1}\left(\boldsymbol{x}-\boldsymbol{k}\right),\\
\widetilde{z}&=\bar{z}
\end{split}
\end{align}
where $P$ is the same as above and $\boldsymbol{k}=\left(k_1,k_2,k_3\right)$ is a constant vector soon to be defined case by case, we put system (\ref{ai}) into a much simpler form,
\begin{align}
\begin{split}
\label{ai0}
\widetilde{\varphi}_1(\widetilde{x})\frac{\partial a_1}{\partial\widetilde{x}}+\widetilde{\varphi}_2(\widetilde{y})\frac{\partial a_1}{\partial\widetilde{y}}+\widetilde{\varphi}_3(\widetilde{z})\frac{\partial a_1}{\partial\widetilde{z}}&=\left(c-h_{11}\right)a_1,\\
\widetilde{\varphi}_1(\widetilde{x})\frac{\partial a_2}{\partial\widetilde{x}}+\widetilde{\varphi}_2(\widetilde{y})\frac{\partial a_2}{\partial\widetilde{y}}+\widetilde{\varphi}_3(\widetilde{z})\frac{\partial a_2}{\partial\widetilde{z}}&=-h,\\
\widetilde{\varphi}_1(\widetilde{x})\frac{\partial a_3}{\partial\widetilde{x}}+\widetilde{\varphi}_2(\widetilde{y})\frac{\partial a_3}{\partial\widetilde{y}}+\widetilde{\varphi}_3(\widetilde{z})\frac{\partial a_3}{\partial\widetilde{z}}&=\left(c-h_{11}\right)a_3.
\end{split}
\end{align}
The solution of this system can be found easily. Notice that $\widetilde{x},\widetilde{y},\widetilde{z}$ describe cylindrical coordinates in the $\bar{x}\bar{y}\bar{z}$-space, which in turn is a linear transformation of the original space.

Summing up, transformation (\ref{tranA}) uncouples the equations of the system, the same way (\ref{tranx}) yields its characteristics. Under these two changes of variables we arrive at system (\ref{ai0}) and then solve it. After the solution is found, the inverse transformations of (\ref{tranA}) and (\ref{tranx}) give the solution of the original system (\ref{Ai}). Although this treatment is general, we distinguish four characteristic cases, depending on the form of the matrix $P$ and the vector $\boldsymbol{k}$, as well as a last and more trivial one, where no transformation is necessary at all.

\subsection{The case $h_{11}\neq0$ and $h_{23}~\text{or}~h_{31}\neq0$}
In this general case the eigenvectors of $Q$ can define the matrix
\begin{equation}
\label{P1}
P=\frac{1}{h\left(h_{31}^2+h_{23}^2\right)^{\frac{1}{2}}}\left(\begin{matrix}
h_{12}h_{23}&-hh_{31}&h_{23}\left(h_{31}^2+h_{23}^2\right)^{\frac{1}{2}}\\
h_{12}h_{31}&hh_{23}&h_{31}\left(h_{31}^2+h_{23}^2\right)^{\frac{1}{2}}\\
-\left(h_{31}^2+h_{23}^2\right)&0&h_{12}\left(h_{31}^2+h_{23}^2\right)^{\frac{1}{2}}
\end{matrix}\right),
\end{equation}
which for this particular choice is orthogonal, $P^{-1}=P^T$, while $\boldsymbol{k}=-H^{-1}\boldsymbol{h}$, making (\ref{char-eq}) homogeneous in terms of $\bar{x},\bar{y},\bar{z}$. So, the new independent variables, according to (\ref{tranx}), are
\begin{align}
\label{tranx1}
\nonumber\widetilde{x}&=\sqrt{(x-k_1)^2+(y-k_2)^2+(z-k_3)^2-\widetilde{z}^2}\\
\widetilde{y}&=\arctan\!\left(\frac{h\left[h_{23}\left(y-k_2\right)-h_{31}\left(x-k_1\right)\right]}{h_{12}h_{23}\left(x-k_1\right)+h_{12}h_{31}\left(y-k_2\right)-\left(h_{31}^2+h_{23}^2\right)\left(z-k_3\right)}\right)\\
\nonumber\widetilde{z}&=\frac{1}{h}\left[h_{23}\left(x-k_1\right)+h_{31}\left(y-k_2\right)+h_{12}\left(z-k_3\right)\right]
\end{align}
Together with (\ref{tranA}), they transform system (\ref{Ai}) into (\ref{ai0}) for $\widetilde{\varphi}_1(\widetilde{x})=h_{11}\widetilde{x},\widetilde{\varphi}_2(\widetilde{y})=-h,\widetilde{\varphi}_3(\widetilde{z})=h_{11}\widetilde{z}$. Its solution is now easily found to be
\begin{align}
\label{sola1}
a_1=\widetilde{z}^{\frac{c}{h_{11}}-1}\widetilde{F}_1\left(c_1,c_2\right),~~~~a_2=\widetilde{y}+\widetilde{F}_2\left(c_1,c_2\right),~~~~a_3=\widetilde{z}^{\frac{c}{h_{11}}-1}\widetilde{F}_3\left(c_1,c_2\right),
\end{align}
where $\widetilde{F}_1,\widetilde{F}_2,\widetilde{F}_3$ are arbitrary functions of the characteristics
\begin{align}
\label{char1}
\begin{split}
c_1&=\frac{\widetilde{z}}{\widetilde{x}},\\
c_2&=h_{11}\widetilde{y}+h\ln\widetilde{z}.
\end{split}
\end{align}

Taking the inverse transformation of (\ref{tranA}), the previous solution can be expressed back in the original components of the vector potential, arriving at
\begin{equation}
\label{A1}
\boldsymbol{A}=\widetilde{z}^{\frac{c}{h_{11}}-1}PR\left(\widetilde{y}\right)\boldsymbol{F}\left(c_1,c_2\right),
\end{equation}
where $\boldsymbol{F}=\left(F_1,F_2,F_3\right)$ is an arbitrary vector function of $c_1$ and $c_2$, coming from $\widetilde{\boldsymbol{F}}=(\widetilde{F}_1,\widetilde{F}_2,\widetilde{F}_3)$, and $R$ is the rotation matrix around the third axis,
\begin{equation}
\label{R}
R(\vartheta)=\left(\begin{matrix}
\cos\vartheta&-\sin\vartheta&0\\
\sin\vartheta&\cos\vartheta&0\\
0&0&1
\end{matrix}\right).
\end{equation}
In terms of the original coordinates, solution (\ref{A1}) is further reduced, though some of the transformed variables are still kept for simplicity,
\begin{align}
\begin{split}
\label{fA1}
A_1&=\widetilde{z}^{\frac{c}{h_{11}}-2}\left\{\left[h\left(x-k_1\right)-h_{23}\widetilde{z}\right]F_1\left(c_1,c_2\right)+\left[h_{31}\left(z-k_3\right)-h_{12}\left(y-k_2\right)\right]F_2\left(c_1,c_2\right)+h_{23}\widetilde{z}F_3\left(c_1,c_2\right)\right\}\\
A_2&=\widetilde{z}^{\frac{c}{h_{11}}-2}\left\{\left[h\left(y-k_2\right)-h_{31}\widetilde{z}\right]F_1\left(c_1,c_2\right)+\left[h_{12}\left(x-k_1\right)-h_{23}\left(z-k_3\right)\right]F_2\left(c_1,c_2\right)+h_{31}\widetilde{z}F_3\left(c_1,c_2\right)\right\}\\
A_3&=\widetilde{z}^{\frac{c}{h_{11}}-2}\left\{\left[h\left(z-k_3\right)-h_{12}\widetilde{z}\right]F_1\left(c_1,c_2\right)+\left[h_{23}\left(y-k_2\right)-h_{31}\left(x-k_1\right)\right]F_2\left(c_1,c_2\right)+h_{12}\widetilde{z}F_3\left(c_1,c_2\right)\right\}
\end{split}
\end{align}

Equation (\ref{Ph}) on the other hand is very similar to either one of system (\ref{ai0}), and can be solved likewise. So, the scalar potential is
\begin{equation}
\label{fPh1}
\Phi=\begin{cases}
\widetilde{z}^{\,2\left(\frac{c}{h_{11}}-1\right)}G\!\left(c_1,c_2\right),&c\neq h_{11}\\
\frac{k}{h}\,\widetilde{y}+G\!\left(c_1,c_2\right),&c=h_{11}
\end{cases}
\end{equation}
up to some additive constant, where $G$ is arbitrary.

Since $k_i$s are quite complicated expressions of $h_i$s, it is preferable in this case for practical purposes to consider the latter in terms of the former, and thus, to conclude that the potentials of the form (\ref{fA1})-(\ref{fPh1}) yield symmetry (\ref{sym1}) for $h_i=-h_{ij}k_j$.

\subsection{The case $h_{11}=0$ and $h_{23}~\text{or}~h_{31}\neq0$}
Here we may use again the matrix $P$ given in (\ref{P1}), but, since $H$ is no longer invertible, we define $\boldsymbol{k}=\frac{1}{h^2}H\boldsymbol{h}$. The new coordinates resemble the ones of the previous case,
\begin{align}
\label{tranx2}
\nonumber\widetilde{x}&=\sqrt{(x-k_1)^2+(y-k_2)^2+(z-k_3)^2-\widetilde{z}^2}\\
\widetilde{y}&=\arctan\!\left(\frac{h\left[h_{23}\left(y-k_2\right)-h_{31}\left(x-k_1\right)\right]}{h_{12}h_{23}\left(x-k_1\right)+h_{12}h_{31}\left(y-k_2\right)-\left(h_{31}^2+h_{23}^2\right)\left(z-k_3\right)}\right)\\
\nonumber\widetilde{z}&=\frac{1}{h}\left(h_{23}x+h_{31}y+h_{12}z\right),
\end{align}
where $k_1=\left(h_{12}h_2-h_{31}h_3\right)/h^2,k_2=\left(h_{23}h_3-h_{12}h_1\right)/h^2,k_3=\left(h_{31}h_1-h_{23}h_2\right)/h^2$, but now $\widetilde{\varphi}_1(\widetilde{x})=0,\widetilde{\varphi}_2(\widetilde{y})=-h,\widetilde{\varphi}_3(\widetilde{z})=\bar{h}_3$, where $\bar{h}_3=\left(h_{23}h_1+h_{31}h_2+h_{12}h_3\right)/h$. So,
\begin{align}
\label{sola2}
a_1=e^{-\frac{c}{h}\widetilde{y}}\widetilde{F}_1\left(c_3,c_4\right),~~~~a_2=\widetilde{y}+\widetilde{F}_2\left(c_3,c_4\right),~~~~a_3=e^{-\frac{c}{h}\widetilde{y}}\widetilde{F}_3\left(c_3,c_4\right),
\end{align}
where the arbitrary functions $\widetilde{F}_1,\widetilde{F}_2,\widetilde{F}_3$ depend now on the characteristics
\begin{align}
\label{char2}
\begin{split}
c_3&=\widetilde{x},\\
c_4&=\bar{h}_3\widetilde{y}+h\widetilde{z},
\end{split}
\end{align}
that actually being the essential difference with the first case.

Thus, the vector potential in the original reference frame is
\begin{equation}
\label{A2}
\boldsymbol{A}=e^{-\frac{c}{h}\widetilde{y}}PR\left(\widetilde{y}\right)\boldsymbol{F}\left(c_3,c_4\right),
\end{equation}
or more simply
\begin{align}
\begin{split}
\label{fA2}
A_1&=e^{-\frac{c}{h}\widetilde{y}}\left\{\left[h\left(x-k_1\right)-h_{23}\widetilde{z}\right]F_1\left(c_3,c_4\right)+\left[h_{31}\left(z-k_3\right)-h_{12}\left(y-k_2\right)\right]F_2\left(c_3,c_4\right)+h_{23}F_3\left(c_3,c_4\right)\right\}\\
A_2&=e^{-\frac{c}{h}\widetilde{y}}\left\{\left[h\left(y-k_2\right)-h_{31}\widetilde{z}\right]F_1\left(c_3,c_4\right)+\left[h_{12}\left(x-k_1\right)-h_{23}\left(z-k_3\right)\right]F_2\left(c_3,c_4\right)+h_{31}F_3\left(c_3,c_4\right)\right\}\\
A_3&=e^{-\frac{c}{h}\widetilde{y}}\left\{\left[h\left(z-k_3\right)-h_{12}\widetilde{z}\right]F_1\left(c_3,c_4\right)+\left[h_{23}\left(y-k_2\right)-h_{31}\left(x-k_1\right)\right]F_2\left(c_3,c_4\right)+h_{12}F_3\left(c_3,c_4\right)\right\}
\end{split}
\end{align}
and the scalar one,
\begin{equation}
\label{fPh2}
\Phi=\begin{cases}
e^{-\frac{2c}{h}\widetilde{y}}\,G\!\left(c_3,c_4\right),&c\neq0\\
\frac{k}{h}\,\widetilde{y}+G\!\left(c_3,c_4\right),&c=0
\end{cases},
\end{equation}
where $\boldsymbol{F}$ and $G$ are now arbitrary functions of $c_3$ and $c_4$.

\subsection{The case $h_{11}\neq0$ and $h_{23}=h_{31}=0$}
Now the third equation of system (\ref{Ai}) is already isolated and so it can be solved independently, yielding $A_3$ directly. This means that simply $P=I$, which is actually the only difference with the first case. All the results obtained there can be reproduced here by simply making this substitution. In particular, setting again $\boldsymbol{k}=-H^{-1}\boldsymbol{h}$, the form of system (\ref{ai0}) in the new variables
\begin{align}
\begin{split}
\label{tranx3}
\widetilde{x}&=\sqrt{\left(x-k_1\right)^2+\left(y-k_2\right)^2}\\
\widetilde{y}&=\arctan\!\left(\frac{y-k_2}{x-k_1}\right)\\
\widetilde{z}&=z-k_3
\end{split}
\end{align}
and consequently its solution remain the same. Therefrom, we end up with the following expression for the vector potential
\begin{equation}
\label{A3}
\boldsymbol{A}=\widetilde{z}^{\frac{c}{h_{11}}-1}R\left(\widetilde{y}\right)\boldsymbol{F}\left(c_1,c_2\right),
\end{equation}
where the arbitrary function $\boldsymbol{F}$ depends again on $c_1$ and $c_2$, but for the variables $\widetilde{x},\widetilde{y},\widetilde{z}$ defined above, noting that $h=h_{12}$ in this case. Thus, in terms of the original coordinates the above solution is written as
\begin{align}
\begin{split}
\label{fA3}
A_1&=\left(z-k_3\right)^{\frac{c}{h_{11}}-2}\left[\left(x-k_1\right)F_1\left(c_1,c_2\right)-\left(y-k_2\right)F_2\left(c_1,c_2\right)\right]\\
A_2&=\left(z-k_3\right)^{\frac{c}{h_{11}}-2}\left[\left(y-k_2\right)F_1\left(c_1,c_2\right)+\left(x-k_1\right)F_2\left(c_1,c_2\right)\right]\\
A_3&=\left(z-k_3\right)^{\frac{c}{h_{11}}-1}F_3\left(c_1,c_2\right),
\end{split}
\end{align}
where
\begin{align}
\label{char3}
\begin{split}
c_1&=\frac{z-k_3}{\sqrt{\left(x-k_1\right)^2+\left(y-k_2\right)^2}},\\
c_2&=h_{11}\arctan\!\left(\frac{y-k_2}{x-k_1}\right)+h_{12}\ln\left(z-k_3\right),
\end{split}
\end{align}
and $k_1=\left(h_{12}h_2-h_{11}h_1\right)/\left(h_{11}^2+h_{12}^2\right),k_2=-\left(h_{12}h_1+h_{11}h_2\right)/\left(h_{11}^2+h_{12}^2\right),k_3=-h_3/h_{11}$. The scalar potential accordingly is
\begin{equation}
\label{fPh3}
\Phi=\begin{cases}
\left(z-k_3\right)^{\,2\left(\frac{c}{h_{11}}-1\right)}G\!\left(c_1,c_2\right),&c\neq h_{11}\\
-\frac{k}{h_{11}}\ln\left(z-k_3\right)+G\!\left(c_1,c_2\right),&c=h_{11}
\end{cases},
\end{equation}
where $G$ is again arbitrary in its arguments.

When $h_{12}=0$, then the characteristic $c_2$ is just $\widetilde{y}$. Thus, solution (\ref{A3}) simply reduces to $\boldsymbol{A}=\left(z-k_3\right)^{\frac{c}{h_{11}}-1}\boldsymbol{F}\!\left(c_1,\widetilde{y}\right)$. In other words, there is no need for transformation (\ref{tranA}) or (\ref{tranx}) at all, because, as we can see from (\ref{Ai}), the system is completely uncoupled.

\subsection{The case $h_{11}=h_{23}=h_{31}=0,h_{12}\neq0$}
This is actually a combination of the last two cases. The new variables are now defined for $P=I$, as in the previous case, but $\boldsymbol{k}=\frac{1}{h^2}H\boldsymbol{h}$ as in the second case, meaning
\begin{align}
\begin{split}
\label{tranx4}
\widetilde{x}&=\sqrt{\left(x-k_1\right)^2+\left(y-k_2\right)^2}\\
\widetilde{y}&=\arctan\!\left(\frac{y-k_2}{x-k_1}\right)\\
\widetilde{z}&=z,
\end{split}
\end{align}
where $k_1=h_2/h_{12},k_2=-h_1/h_{12}$ and $k_3=0$. The form of $\widetilde{\varphi_i}$ in terms of the above variables is also the same as is in the second case for $h=h_{12}$ and $\bar{h}_3=h_3$. Thus, the vector potential is
\begin{equation}
\label{A4}
\boldsymbol{A}=e^{-\frac{c}{h_{12}}\widetilde{y}}R\left(\widetilde{y}\right)\boldsymbol{F}\left(c_3,c_4\right),
\end{equation}
or
\begin{align}
\begin{split}
\label{fA4}
A_1&=e^{-\frac{c}{h_{12}}\arctan\!\left(\frac{y-k_2}{x-k_1}\right)}\left[\left(x-k_1\right)F_1\left(c_3,c_4\right)-\left(y-k_2\right)F_2\left(c_3,c_4\right)\right]\\
A_2&=e^{-\frac{c}{h_{12}}\arctan\!\left(\frac{y-k_2}{x-k_1}\right)}\left[\left(y-k_2\right)F_1\left(c_3,c_4\right)+\left(x-k_1\right)F_2\left(c_3,c_4\right)\right]\\
A_3&=e^{-\frac{c}{h_{12}}\arctan\!\left(\frac{y-k_2}{x-k_1}\right)}F_3\left(c_3,c_4\right)
\end{split}
\end{align}
and the scalar one,
\begin{equation}
\label{fPh4}
\Phi=\begin{cases}
e^{-\frac{2c}{h_{12}}\arctan\!\left(\frac{y-k_2}{x-k_1}\right)}G\!\left(c_3,c_4\right),&c\neq0\\
\frac{k}{h_{12}}\arctan\!\left(\frac{y-k_2}{x-k_1}\right)+G\!\left(c_3,c_4\right),&c=0
\end{cases},
\end{equation}
where functions $\boldsymbol{F}$ and $G$ are arbitrary in the indicated arguments. The latter are defined again through (\ref{char2}), but for the variables introduced in this case, meaning
\begin{align}
\label{char4}
\begin{split}
c_3&=\sqrt{\left(x-k_1\right)^2+\left(y-k_2\right)^2},\\
c_4&=h_3\arctan\!\left(\frac{y-k_2}{x-k_1}\right)+h_{12}z.
\end{split}
\end{align} 

\subsection{The case $h_{ij}=0,h_1\neq0$}
Now no transformation is needed at all. Equations (\ref{Ai}) are uncoupled, and each one of them can be solved independently. The solution for the vector potential is
\begin{equation}
\label{fA5}
\boldsymbol{A}=e^{\frac{c}{h_1}x}\boldsymbol{F}\!\left(h_2x-h_1y,h_3x-h_1z\right),
\end{equation}
while for the scalar
\begin{equation}
\label{fPh5}
\Phi=\begin{cases}
e^{\frac{2c}{h_1}x}G\!\left(h_2x-h_1y,h_3x-h_1z\right),&c\neq0\\
-\frac{k}{h_1}\,x+G\!\left(h_2x-h_1y,h_3x-h_1z\right),&c=0
\end{cases}
\end{equation}
for some arbitrary vector function $\boldsymbol{F}$ and a scalar one $G$.

\begin{rem}
In all of the above cases, the constants $h_{ij},h_i\text{ (or }k_i\text{ in the first case)}$ and $c$, that appear in the symmetry $\text{v}_{\text{L}}$, are also present in the expressions for $\boldsymbol{A}$ and $\Phi$. In this way, they define both the electromagnetic field and at the same time the symmetry admitted by system (\ref{lor}). In other words, the cases presented in this section represent families of potentials, which, apart from time translations, are compatible with a second Lie point symmetry of the form (\ref{sym1}). So, $\text{v}_{\text{L}}$ is actually one symmetry, and along with $\partial_t$ they generate two-dimensional symmetry algebras for each case, respectively. Further restrictions on the arbitrariness of $\boldsymbol{A}$ and $\Phi$ can yield symmetry algebras of higher dimensions, but these cases will not be considered here.
\end{rem}

\section{Noether point symmetries}
In retrospect, the introduction of the potentials of the electromagnetic field leads immediately to a different viewpoint regarding equations (\ref{lor}). They form an Euler-Lagrange system, stemming from the function
\begin{equation}
\label{L}
L\left(\boldsymbol{x},\dot{\boldsymbol{x}}\right)=\frac{1}{2}\,\dot{\boldsymbol{x}}^2+\dot{\boldsymbol{x}}\cdot\boldsymbol{A}(\boldsymbol{x})-\Phi(\boldsymbol{x}).
\end{equation}
Subsequently, the question of Noether symmetries, providing us with first integrals of motion, naturally arises. Unlike the case of Lie point ones, now Maxwell's equations are obviously implied from the beginning.

Considering that Noether point symmetries are a subset of the Lie point ones, already found, we do not have to employ a new query from the beginning. We may check instead if and when the general form of the symmetry generator (\ref{sym0}) is of Noether type. Following \cite{sa-ca}, we note that the vector field (\ref{X}) defines a Noether point symmetry, if a function $\widetilde{f}(t,\boldsymbol{x})$ exists, such that
\begin{equation}
\label{nsc}
\xi\frac{\partial L}{\partial t}+\varphi_i\frac{\partial L}{\partial x_i}+\left(\frac{d\varphi_i}{dt}-\frac{d\xi}{dt}\dot{x}_i\right)\frac{\partial L}{\partial \dot{x}_i}+\frac{d\xi}{dt}L=\frac{d\widetilde{f}}{dt}.
\end{equation}

Substituting the time-independent Lagrangian given in (\ref{L}) and the components of the generator from (\ref{sym0}) results in a second-degree polynomial in terms of the velocity components. This equation must hold identically for all $t,x_i,\dot{x}_i$ and so every coefficient of this polynomial must vanish, yielding the following five equations 
\begin{align}
\label{n1}
c&=0,\\
\label{n2}
\left(h_{jk}x_k+h_j\right)\frac{\partial A_i}{\partial x_j}+\frac{2}{3}\,h_{jj}A_i-h_{ij}A_j&=\frac{\partial\widetilde{f}}{\partial x_i},\\
\label{n3}
-\left(h_{jk}x_k+h_j\right)\frac{\partial\Phi}{\partial x_j}-\frac{2}{3}\,h_{jj}\Phi&=\frac{\partial\widetilde{f}}{\partial t}.
\end{align}
The first equation clearly rules the dilation $t\partial_t$ out as a Noether symmetry candidate. On the other hand, the integrability conditions of (\ref{n2}) and (\ref{n3}), $\widetilde{f}_{x_ix_j}=\widetilde{f}_{x_jx_i}$ and $\widetilde{f}_{tx_i}=\widetilde{f}_{x_it}$, which guarantee the existence of $\widetilde{f}$, expressed in terms of $\boldsymbol{B}$ and $\boldsymbol{E}$, lead back to (\ref{B}) and (\ref{E}) again for $c=0$. Furthermore, from (\ref{n2}) and (\ref{n3}), we can easily deduce that $\widetilde{f}(t,\boldsymbol{x})=kt+f(\boldsymbol{x})$, where $k$ is a constant and $f$ arbitrary. Thus, we recover the restrictions (\ref{AF}) and (\ref{Ph}), for $c=0$, which, going the other way round, were derived, when (\ref{B}) and (\ref{E}) were integrated.

So, the Noether symmetry condition, apart from (\ref{n1}), does not involve any other constraint. Of course, the very expression of the Lagrange function requires Maxwell's equations from the start. For a real physical problem, these would be implemented in the Lie symmetry case, as well. On this ground, the conditions placed upon the electromagnetic field, in terms of $\boldsymbol{A}$ and $\Phi$, are the previous ones for $c=0$. Once again the gauge invariance of the vector potential can be used as in section 3, where the restriction (\ref{g}) for $c=0$ now reduces to $\boldsymbol{\varphi}\cdot\nabla g=f$. In conclusion, equations (\ref{lor}) admit Noether point symmetries of the general form
\begin{align}
\begin{split}
\label{nsym0}
\xi&=2h_{11}t+h_0,\\
\varphi_1&=h_{11}x+h_{12}y-h_{31}z+h_1,\\
\varphi_2&=-h_{12}x+h_{11}y+h_{23}z+h_2,\\
\varphi_3&=h_{31}x-h_{23}y+h_{11}z+h_3
\end{split}
\end{align}
for electromagnetic fields coming from the potentials described earlier in the previous section, by setting $c=0$.

We emphasize again that only $h_0$ is completely arbitrary, corresponding to the symmetry $\partial_t$. The rest of the constants define a second Noether symmetry,
\begin{align}
\label{nsym1}
\text{v}_{\text{N}}=&\,2h_{11}t\frac{\partial}{\partial t}+\varphi_1\frac{\partial}{\partial x}+
\varphi_2\frac{\partial}{\partial y}+\varphi_3\frac{\partial}{\partial z},
\end{align}
which the system has for a particular form of the electromagnetic field, specified again by them. While the invariant that corresponds to the first symmetry $\partial_t$ is the well known Hamiltonian of the system, a second integral of motion may arise from $\text{v}_{\text{N}}$.

\section{Integrals of motion}
According to \cite{sa-ca}, the integral of motion, which via Noether's theorem corresponds to a symmetry of (\ref{L}), is given by the following equation:
\begin{equation}
I=\xi L+\left(\varphi_i-\xi\dot{x}_i\right)\frac{\partial L}{\partial\dot{x}_i}-\widetilde{f}.
\end{equation}
Substitution of the Lagrangian of the system yields
\begin{equation}
\label{I0}
I=\boldsymbol{\varphi}\cdot\left(\dot{\boldsymbol{x}}+\boldsymbol{A}\right)-\xi\left(\frac{1}{2}\,\dot{\boldsymbol{x}}^2+\Phi\right)-\widetilde{f}.
\end{equation}

Equations (\ref{n2}) and (\ref{n3}) for $\varphi_i=0$ and $\xi=1$ trivially result in a constant function $\widetilde{f}$. Thus, from the symmetry $\partial_t$, we recover through (\ref{I0}) the integral
\begin{equation}
H\!\left(\boldsymbol{x},\dot{\boldsymbol{x}}\right)=\frac{1}{2}\,\dot{\boldsymbol{x}}^2+\Phi\!\left(\boldsymbol{x}\right),
\end{equation}
which is commonly used as the Hamiltonian function of the system, expressing the particle's energy.

Before finding the integral that corresponds to the second symmetry (\ref{nsym1}), we comment that a second constant of motion would be of real value, if it is functionally independent of the already known Hamiltonian and in involution with it with respect to the corresponding Poisson bracket. For this to be the case, we set $h_{11}=0$, and also require $\widetilde{f}$ to be time-independent, that is, $k=0$. Then, using the gauge transformation $\boldsymbol{A}\longrightarrow\boldsymbol{A}+\nabla g$ of section 3, $\widetilde{f}=f$ will not enter at all in (\ref{I0}), since equation (\ref{g}) for $c=0$ yields $\boldsymbol{\varphi}\cdot\nabla g=f$. Thus, we restrict our attention to a linear integral,
\begin{equation}
\label{I}
I\left(\boldsymbol{x},\dot{\boldsymbol{x}}\right)=\boldsymbol{\varphi}\cdot\left(\dot{\boldsymbol{x}}+\boldsymbol{A}\right),
\end{equation}
either in terms of the velocities or the canonical momentums $p_i=L_{\dot{x}_i}=\dot{x}_i+A_i$. This is actually a generalization of the relevant two-dimensional results in \cite{dor} and \cite{hiet}, and not included in Lewis's search for quadratic invariants \cite{lewis}. It is worth noticing that in this way the scalar potential defines the form of the Hamiltonian, while the vector potential the form of the second integral.

We distinguish the following three cases, which correspond to the second, fourth and fifth one described in section 3 for $c=k=0$.

\subsection{The case $h_{23}~\text{or}~h_{31}\neq0$}
The vector and scalar potentials in this case are given in (\ref{fA2}) and (\ref{fPh2}), respectively, for $c=k=0$. The corresponding integral is
\begin{align}
\label{I1}
\nonumber I=\left(h_{12}y-h_{31}z+h_1\right)\dot{x}+\left(-h_{12}x+h_{23}z+h_2\right)\dot{y}+\left(h_{31}x-h_{23}y+h_3\right)\dot{z}\\
-\,h^2\widetilde{x}^2F_2\!\left(\widetilde{x},\bar{h}_3\widetilde{y}+h\widetilde{z}\right)+h\bar{h}_3F_3\!\left(\widetilde{x},\bar{h}_3\widetilde{y}+h\widetilde{z}\right)
\end{align}
where $\widetilde{x},\widetilde{y},\widetilde{z}$ are given in (\ref{tranx2}).

\subsection{The case $h_{23}=h_{31}=0$}
The vector and scalar potentials in this case are given in (\ref{fA4}) and (\ref{fPh4}), respectively, for $c=k=0$. The corresponding integral is
\begin{equation}
\label{I2}
I=\left(h_{12}y+h_1\right)\dot{x}+\left(-h_{12}x+h_2\right)\dot{y}+h_3\dot{z}-h_{12}\widetilde{x}^2F_2\!\left(\widetilde{x},h_3\widetilde{y}+h_{12}\widetilde{z}\right)+h_3F_3\!\left(\widetilde{x},h_3\widetilde{y}+h_{12}\widetilde{z}\right)
\end{equation}
where $\widetilde{x},\widetilde{y},\widetilde{z}$ are given in (\ref{tranx4}).

\subsection{The case $h_{ij}=0,h_1\neq0$}
The vector and scalar potentials in this case are given in (\ref{fA5}) and (\ref{fPh5}), respectively, for $c=k=0$. The corresponding integral is
\begin{align}
\label{I3}
\nonumber I=h_1\dot{x}+h_2\dot{y}+h_3\dot{z}+h_1F_1\!\left(h_2x-h_1y,h_3x-h_1z\right)+h_2F_2\!\left(h_2x-h_1y,h_3x-h_1z\right)\\
+\,h_3F_3\!\left(h_2x-h_1y,h_3x-h_1z\right)
\end{align}
\begin{rem}
In all of the above cases, the scalar potential reduces to an arbitrary function of the related characteristics. For each case, there also exists a suitable coordinate system, where the vector potential is an arbitrary function of the corresponding characteristics, as well. 
\end{rem}

There is another justification for focusing on these types of integrals, which can be apparent when investigating the dynamics of the magnetic field itself. The latter offer a better insight into the magnetic fields found in section 3.

\section{Magnetic field lines}
In this section, we compare, in terms of symmetries, system (\ref{lor}), which describes the particle's orbit, with the system of equations
\begin{equation}
\label{mag}
\frac{d\boldsymbol{x}}{d\tau}=\boldsymbol{B}(\boldsymbol{x}),
\end{equation}
which describes the integral curves of the magnetic field $\boldsymbol{B}$, commonly known as magnetic field lines. The independent variable here, denoted by $\tau$, is related to the line element of these curves. From the physical point of view, since $t$ and $\tau$ carry very different meanings, such a comparison can only be made on the ground of spatial symmetries independent of them, i.e. generated by a vector field of the form $X=\varphi_{i}(\boldsymbol{x})\partial_{x_i}$.

For system (\ref{lor}), these can be recovered from section 2 for $\xi=0$, meaning $c=2h_{11}$ and $h_0=0$. So, in this case, the coefficients $\varphi_i$ are restricted to the form (\ref{g2}) together with (\ref{g4}) and (\ref{g5}), while conditions (\ref{B}) for the magnetic field are
\begin{equation}
\label{B1}
\varphi_j\frac{\partial B_i}{\partial x_j}=-h_{jj}B_i+\frac{\partial\varphi_i}{\partial x_j}B_j.
\end{equation}

On the other hand, by prolonging $X$ up now to the first order derivatives $x_i'(\tau)$, the symmetry condition for system (\ref{mag})
\begin{equation}
X^{(1)}\!\left(x_i'-B_i\right)=0
\end{equation}
on its solutions leads simply to
\begin{equation}
\label{B2}
\varphi_j\frac{\partial B_i}{\partial x_j}=\frac{\partial\varphi_i}{\partial x_j}B_j,
\end{equation}
without predefining the form of the symmetries.

From the above we conclude that, for a given magnetic field, which satisfies equations (\ref{B1}), the symmetries of (\ref{lor}) cannot meet the requirements (\ref{B2}) for system (\ref{mag}) when $h_{11}\neq0$. All the same, for $h_{11}=0$, where conditions (\ref{B1}) and (\ref{B2}) are identical, system (\ref{mag}) may have more symmetries than the linear ones of system (\ref{lor}). Thus, in general, not every symmetry of the charged particle motion is a symmetry of the magnetic field lines and vice versa. If, however, we limit our choices to symmetries of the form
\begin{equation}
\label{nsym2}
\text{v}=\left(h_{12}y-h_{31}z+h_1\right)\frac{\partial}{\partial x}+\left(-h_{12}x+h_{23}z+h_2\right)\frac{\partial}{\partial y}+\left(h_{31}x-h_{23}y+h_3\right)\frac{\partial}{\partial z},
\end{equation}
then these are preserved by both systems for the same magnetic field $\boldsymbol{B}$.

Actually, we have already encountered $\text{v}$: it is a Noether symmetry of (\ref{lor}) that corresponds to the integral (\ref{I}) described in the previous section. However, (\ref{mag}) is an Euler-Lagrange system, too, for the Lagrangian function
\begin{equation}
\label{Lmag}
L_m\!\left(\boldsymbol{x},\boldsymbol{x}'\right)=\boldsymbol{x}'\cdot\boldsymbol{A}(\boldsymbol{x}).
\end{equation}
The Noether symmetry condition for these equations,
\begin{equation}
\varphi_i\frac{\partial L_m}{\partial x_i}+\frac{d\varphi_i}{d\tau}\frac{\partial L_m}{\partial x'_i}=\frac{d\widetilde{f}}{d\tau},
\end{equation}
is also satisfied by $\text{v}$, giving $\widetilde{f}_\tau=0$ and the previous constrains on the vector potential, i.e. equations (\ref{AF}) for $c=h_{11}=0$. Using the same gauge transformation to determine $\boldsymbol{A}$, the corresponding integral, $\varphi_i L_{x_i'}-\widetilde{f}$, for system (\ref{mag}) becomes
\begin{equation}
\label{Imag}
I_m=\boldsymbol{\varphi}\cdot\boldsymbol{A}.
\end{equation}
This is a projection of the integral (\ref{I}) on the original configuration space $\mathbb{R}^3$, and therefore, according to the form of the symmetry and the vector potential, it, too, is seperated into three cases, which are the ones of the previous section without the velocities. 

In conclusion, whenever the motion of the charged particle is confined in the hypersurface $\boldsymbol{\varphi}\cdot\left(\dot{\boldsymbol{x}}+\boldsymbol{A}\right)=\text{const.}$, the magnetic field lines lie on the surface $\boldsymbol{\varphi}\cdot\boldsymbol{A}=\text{const.}$. In this case, both systems enjoy a symmetry of the form (\ref{nsym2}).

\begin{rem}
The system of the magnetic field lines is integrable, when it admits the above symmetry $\text{v}$, as shown, for example, in \cite{gas}. Its Hamiltonian formulation in this case has also been studied in \cite{me-wi,cas}. Further results on the reduction of divergence-free vector fields, like the magnetic field, with divergence-free symmetries, like the above, can be found in \cite{ha-me} for the three-dimensional case and in \cite{huan,zhen} for the $n$-dimensional one.
\end{rem}

\section{Conclusions}

We have found five classes of stationary electromagnetic fields in terms of the potentials, which yield a second Lie point symmetry, besides time translations, for the three-dimensional autonomous nonrelativistic charged particle motion. The analysis showed that in the case of non-homogeneous arbitrary fields the only possible symmetries are linear, consisting of dilations, rotations and translations. Besides time dilations, they have all proven to be of Noether type, too, without any further restrictions on the potentials of the electromagnetic field. The corresponding invariants have been constructed, in particular focusing on three cases, where the integrals are functionally independent of the Hamiltonian and in involution with it. Thus, a total reduction of four can be made, and then further investigate the system in only two variables.

Another aspect of these three cases is that the same symmetry is also admitted by the system of the magnetic field lines, yielding an integral of motion for the latter, too. Time-independent magnetic fields lying on a surface, with some geometrical (usually axial or helical) symmetry, are very often used to describe the equilibrium state of plasma configurations in the context of ideal magnetohydrodynamics. The results obtained in this work may be useful for the determination of such symmetric magnetic surfaces, by finding new solutions of the Grad-Shafranov equation, for example. In addition, the comparison between the two integrals, (\ref{I}) and (\ref{Imag}), could also relate the behaviour, and possibly the confinement, of the charged particle to the dynamics of the magnetic field. This relation could be further analysed, after reduction of the particle's trajectories and integration of the magnetic field lines.

A desirable development of the present work would be the treatment of the non-autonomous case, where there are no known integrals in general. Another possible extension is the investigation of complete integrability of system (\ref{lor}), i.e. considering subcases of those presented in section 3, admitting more than one symmetry.

\subsection*{Acknowledgements}
The authors would like to thank D. Tsoubelis, L. Vlahos and H. Isliker for carefully reading the manuscript and valuable suggestions, as well as P. G. L. Leach and G. Papamikos for useful comments and helpful discussions. N. K. acknowledges support from the European Fusion Programme (Association EURATOM-Hellenic Republic) and the Hellenic Secretariat of Research and Technology. The sponsors do not bear any responsibility for the contents of this work.


\end{document}